\documentclass[12pt]{article}
\usepackage{color}
\usepackage{stmaryrd}
\usepackage{bbm}
\usepackage{mathrsfs}
\usepackage{amsfonts}
\usepackage{amssymb}
\usepackage{amsmath}
\usepackage{amsthm}
\usepackage{cases}
\usepackage{indentfirst}
\usepackage{lscape}
\usepackage{graphicx}
\usepackage[sort&compress,numbers]{natbib}

\allowdisplaybreaks[4]

\def\be{\begin{eqnarray}} \def\ee{\end{eqnarray}} 
  \def\({\left(} \def\){\right)}
\def\bc{\begin{center}} 
\def\ec{\end{center}}  
\def\bey{\begin{eqnarray*}}\def\eey{\end{eqnarray*}}

\begin{document}

\title{ {\bf  A generalized super integrable hierarchy of Dirac type}
\footnotetext{*Corresponding author, Email:  houzili187@163.com }}
\author{Yujian Ye$^a$, \quad  Zhihui Li$^{b, *}$,  \quad Shoufeng Shen$^b$, \quad Chunxia Li$^c$
}
\date{}
\maketitle
\begin{center}
\begin{minipage}{135mm}
\noindent {\it \small $^{a}$ School of Management,
Hangzhou Dianzi University, Hangzhou 310018, China\\ $^b$  Department of Applied Mathematics, Zhejiang
University of Technology, Hangzhou 310023, China\\$^c$  School of Mathematical Sciences, Capital Normal University, Beijing 100048, China}\\
\end{minipage}
\end{center}

\begin{abstract}

In this letter, a new generalized matrix spectral problem of Dirac type associated with the super Lie algebra $\mathcal{B}(0,1)$ is proposed and its corresponding super integrable hierarchy is constructed.

\end{abstract}

\vspace{0.5cm}

\noindent{\bf Keywords:}Super Lie algebra; A generalized super Dirac matrix spectral problem; Zero curvature equation; Super integrable hierarchy of Dirac type.

\vspace{0.5cm}

\noindent{\bf PACS numbers:} 02.30.Ik

\noindent{\bf MSC numbers:} 35Q58; 35Q51
\vspace{0.5cm}

 \numberwithin{equation}{section}

\section{Introduction}

Recently, some well-known matrix spectral problems have been generalized to their super counterparts. In literature, some super integrable hierarchies have been constructed successfully, such as the super AKNS hierarchy, the super Dirac hierarchy, the super KdV hierarchy and the super KP hierarchy \cite{ku, mr, lz, lz2, tu, liu, hu, lh, mhq, yhmc, hycz, ma, yhch, tx, gw, gw2, fh, fan, txs, hyh, zhou, xl, zr}.  Meanwhile,  inverse scattering transformation,  B\"{a}cklund transformation, Darboux transformation and Hirota's bilinear method, etc. have also been applied to study super integrable systems.

The well-known Dirac spectral problem reads as \cite{Dirac}
\begin{equation}\label{DSP1}
\phi_x=U\phi,\,\, U=p\epsilon_1+(\lambda+q)\epsilon_2+(-\lambda+q)\epsilon_3
\end{equation}
where $\epsilon_1, \, \epsilon_2,\, \epsilon_3$ are the basis of the algebra $sl(2,R)$ given by
\begin{equation}
\epsilon_1=\left[\begin{array}{cc}1&0\\0&-1\end{array}\right],\, \epsilon_2=\left[\begin{array}{cc}0&1\\0&0\end{array}\right],\, \epsilon_3=\left[\begin{array}{cc}0&0\\1&0\end{array}\right].
\end{equation}
In literature, its corresponding soliton hierarchy and bi-Hamiltonian structure have been established \cite{Dirac1,DirM}.  

Later on, the Dirac spectral problem \eqref{DSP1} associated with $sl(2,R)$ was generalized to the super Dirac spectral problem associated with the super Lie algebra $\mathfrak{B}(0,1)$.
Generally speaking, the super Lie algebra $\mathfrak{B}(0,1)$ has different bases. Here we take the basis $\{e_1, e_2, e_3, e_4, e_5\}$ of $\mathfrak{B}(0,1)$ to be \cite{hu, zr}
\begin{eqnarray*}
& & e_1=\left[\begin{array} {ccc}
1&0 & 0  \cr
0&-1 & 0  \cr
0&0 & 0
\end{array}\right],\qquad e_2=\left[\begin{array} {ccc}
0&1 & 0  \cr
0&0 & 0  \cr
0&0 & 0
\end{array}\right],\qquad e_3=\left[\begin{array} {ccc}
0&0 & 0  \cr
1&0 & 0  \cr
0&0 & 0
\end{array}\right],
\\
& & e_4=\left[\begin{array} {ccc}
0&0 & 1  \cr
0&0 & 0  \cr
0&-1 & 0
\end{array}\right],\qquad e_5=\left[\begin{array} {ccc}
0&0 & 0  \cr
0&0 & 1  \cr
1&0 & 0
\end{array}\right].
\end{eqnarray*}
and the super Dirac matrix spectral problem is defined by
\begin{eqnarray}\label{a2}
&&\phi_x=[p e_1+(\lambda+q) e_2+(-\lambda+q) e_3+\alpha e_4+\beta e_5]\phi\nonumber\\
&&\quad\,\,=\left[\begin{array} {ccc}
p & \lambda+q &\alpha\cr
-\lambda+q & -p &\beta\cr
\beta &-\alpha &0
\end{array}\right]\phi.
\end{eqnarray}
Here we would like to point out that potentials $p, q, \alpha, \beta$ are functions of $x$ and $t$ with different parities. Potentials $p,\, q$  and the spectral parameter $\lambda$ are even elements whose parities are $|p|=|q|=|\lambda|=0$, while potentials $\alpha$ and $\beta$ are odd elements whose parities are $|\alpha|=|\beta|=1$. Physically, $p, q$ and $\lambda$ are bosonic, and $\alpha$ and $\beta$ are fermionic. The super Dirac hierarchy and its super Hamiltonian structure, Bargamann constraint and binary nonlinearization are considered respectively, corresponding to the super matrix spectral problem \eqref{a2} in \cite{mhq, yhmc}. 

In this paper, we will consider the following super matrix spectral problem
\begin{eqnarray}\label{a3}
&&\phi_x=U\phi=[p e_1+(\lambda+q+h) e_2+(-\lambda+q-h) e_3+\alpha e_4+\beta e_5]\phi\\
& & U=\left[\begin{array} {ccc}
p & \lambda+q+h &\alpha\cr
-\lambda+q-h & -p &\beta\cr
\beta &-\alpha &0
\end{array}\right]\phi,
\quad h=c(p^2+q^2+2
\alpha\beta),\nonumber
\end{eqnarray}
where $c$ is an even constant. It is obvious that \eqref{a3} is reduced to the existing super Dirac matrix spectral problem \eqref{a2} by setting $c=0$. This is why we call \eqref{a3} a generalized super Dirac matrix spectral problem. The remainder of this paper is organized as follows. In Section 2, we will establish the corresponding integrable hierarchy of the generalized super Dirac matrix spectral problem \eqref{a3} by the standard procedure. Section 3 is devoted to conclusions and discussions.

\section{Generalized  super integrable  hierarchy of Dirac type}
In this section, we will establish the generalized super integrable hierarchy of Dirac type corresponding to \eqref{a3} by following the standard procedure. 
%
%
%
%
%
%
%
%
%
%
%
%
%
%
%

We will start by solving the stationary zero-curvature equation 
\begin{equation}
W_x=[U,W].
\end{equation}
In order to do so, we assume $W$ in the following form
\begin{eqnarray}\label{b2}
& & W=C e_1+(A+B) e_2+(A-B) e_3+\rho e_4+\sigma e_5=\left[\begin{array} {ccc}
C   &A+B &\rho\cr
A-B & -C &\sigma\cr
\sigma &-\rho&0
\end{array}\right],
\end{eqnarray}
where $|A|=|B|=|C|=0$ and $|\rho|=|\sigma|=1$. Under the assumption, we have 
\begin{equation}\label{b3}
\left\{\begin{array}{l} A_x=-2\lambda C+2pB-\alpha\rho+\beta\sigma-2hC,\vspace{2mm}\\
 B_x=2pA-2qC-\alpha\rho-\beta\sigma,\vspace{2mm}\\
 C_x=2\lambda A-2qB+\alpha\sigma+\beta\rho+2hA,\vspace{2mm}\\
 \rho_x=(\lambda+q)\sigma-\beta(A+B)-\alpha C+p\rho+h\sigma,\vspace{2mm}\\
 \sigma_x=(-\lambda+q)\rho-p\sigma-\alpha(A-B)+\beta C-h\rho.\end{array}\right.
\end{equation}
We expand $W$ in terms of Laurent series 
\begin{eqnarray*}
& & \qquad W=\left[\begin{array} {ccc}
C   & A+B  &\rho   \cr
A-B   & -C &\sigma \cr
\sigma &-\rho &0
\end{array}\right]=\sum_{i=0}^{+\infty}\left[\begin{array} {ccc}
C_i  & A_i+B_i  &\rho_i   \cr
A_i-B_i   & -c_i &\sigma_i \cr
\sigma_i &-\rho_i &0
\end{array}\right]\lambda^{-i}.
\end{eqnarray*}
By comparing the coefficients of the different powers of $\lambda$, we obtain the following recursion relations,
\begin{equation}\label{b4}
\left\{ \begin{array}{l}
A_{i+1}=\frac{1}{2}C_{ix}+qB_i-\frac{1}{2}\alpha\sigma_i-\frac{1}{2}\beta\rho_i-hA_i,\vspace{2mm}\\
C_{i+1}=-\frac{1}{2}A_{ix}+pB_i-\frac{1}{2}\alpha\rho_i+\frac{1}{2}\beta\sigma_i-hC_i\vspace{2mm}\\
\rho_{i+1}=-\sigma_{ix}+q\rho_i-p\sigma_i-\alpha(A_i-B_i)+\beta C_i-h\rho_i\vspace{2mm}\\
\sigma_{i+1}=\rho_{ix}-q\sigma_i-p\rho_i+\beta(A_i+B_i)+\alpha C_i-h\sigma_i\vspace{2mm}\\
B_{i+1,x}=2pA_{i+1}-2qC_{i+1}-\alpha\rho_{i+1}-\beta\sigma_{i+1}
,\end{array}
\right. \qquad i\geq 0.
\end{equation}
Upon chosing the initial value $\{A_0, B_0, C_0, \rho_0, \sigma_0\}$ to be
\begin{eqnarray}\label{b4}
& & A_0=C_0=\rho_0=\sigma_0=0,\qquad B_0=1
\end{eqnarray}
and imposing the integration condition
\begin{eqnarray*}
& & A_i|_{u=0}=B_i|_{u=0}=C_i|_{u=0}=\rho_i|_{u=0}=\sigma_i|_{u=0}=0, \qquad i\geq 1,
\end{eqnarray*}
all the coefficient sets $\{A_i, B_i, C_i, \rho_i, \sigma_i\}$ can be determined uniquely by the recursion relations. The first few sets are given by
\begin{eqnarray*}
& &A_1=q,\qquad B_1=0,\qquad C_1=p,\qquad \rho_1=\alpha,\qquad \sigma_1=\beta;\\
& &A_2=\frac{1}{2}p_x-\epsilon q(p^2+q^2+2\alpha\beta),\\
& &B_2=\frac{1}{2}(p^2+q^2)+\alpha\beta,\\
& &C_2=-\frac{1}{2}q_x-\epsilon p(p^2+q^2+2\alpha\beta),\\
& &\rho_2=-\beta_x-\epsilon(p^2+q^2)\alpha,\\
& &\sigma_2=\alpha_x-\epsilon(p^2+q^2)\beta;\\
& &A_3=-\frac{1}{4}q_{xx}-\frac{1}{2}\alpha\alpha_{x}+\frac{1}{2}\beta\beta_{x}+q\alpha\beta+\frac{1}{2}q(p^2+q^2)
-\frac{1}{2}\epsilon[p(p^2+q^2+2\alpha\beta)_x+2p_x(p^2+q^2+2\alpha\beta)]\\
& &\qquad+\epsilon^2q(p^2+q^2)(p^2+q^2+4\alpha\beta),\\
& &B_3=\frac{1}{2}(p_xq-pq_x)+\alpha\alpha_x+\beta\beta_x-\epsilon(p^2+q^2)(p^2+q^2+4\alpha\beta),\\
& &C_3=-\frac{1}{4}p_{xx}+\frac{1}{2}\alpha\beta_x-\frac{1}{2}\alpha_x\beta+p\alpha\beta+\frac{1}{2}p(p^2+q^2)
+\frac{1}{2}\epsilon[q(p^2+q^2+2\alpha\beta)_x+2q_x(p^2+q^2+2\alpha\beta)]\\
& &\qquad+\epsilon^2p(p^2+q^2)(p^2+q^2+4\alpha\beta),\\
& &\rho_3=-\alpha_{xx}-q\beta_x-p\alpha_x-\frac{1}{2}p_x\alpha-\frac{1}{2}q_x\beta+\frac{1}{2}(p^2+q^2)\alpha
+\epsilon[(p^2+q^2)_x\beta+2(p^2+q^2+\alpha\beta)\beta_x]\\
& &\qquad+\epsilon^2(p^2+q^2)^2\alpha,\\
& &\sigma_3=-\beta_{xx}-q\alpha_x+p\beta_x+\frac{1}{2}p_x\beta-\frac{1}{2}q_x\alpha+\frac{1}{2}(p^2+q^2)\beta
-\epsilon[(p^2+q^2)_x\alpha+2(p^2+q^2+\alpha\beta)\alpha_x]\\
& &\qquad+\epsilon^2(p^2+q^2)^2\beta.
\end{eqnarray*}

Finally, let us the associate the super Dirac matrix spectral problem \eqref{a3} with the following auxiliary spectral problem 
\begin{eqnarray}\label{b5}
& &\phi_{t_m}=V^{[m]}\phi,\\
& &  V^{[m]}=(\lambda^{m}W)_++\Delta_m
=(\lambda^{m}W)_++\left[\begin{array}{ccc}
0  & f_m&0\cr
-f_m & 0 &0 \cr
0 & 0&0
\end{array}\right],
\qquad m\geq 0,\nonumber
\end{eqnarray}
where $P_+$ denotes the polynomial part of $P$ in $\lambda$. The compatibility condition of \eqref{a3} and \eqref{b5} gives the zero curvature equation
\begin{eqnarray*}
& & U_{t_m}-V_x^{[m]}+\left[U,V^{[m]}\right]=0, \qquad m\geq 0
\end{eqnarray*}
which is equivalent to the system
\begin{eqnarray}\label{b7}
\left\{ \begin{array}{l}
p_{t_m}=2A_{m+1}+2qf_m,\\
q_{t_m}=-2C_{m+1}-2pf_m,\\
\alpha_{t_m}=\sigma_{m+1}+\beta f_m,\\
\beta_{t_m}=-\rho_{m+1}-\alpha f_m,\\
{f_m}_x=h_{t_m}.\end{array}
\right.\end{eqnarray}
By some simple calculations, we have
\begin{eqnarray}\label{b8}
& &{f_m}_x=h_{t_m}=\epsilon(2pp_{t_m}+2qq_{t_m}+2\alpha_{t_m}\beta+2\alpha\beta_{t_m})\nonumber\\
& &\qquad=\epsilon\left[2p(2A_{m+1}+2qf_m)+2q(-2C_{m+1}-2pf_m)+2(\sigma_{m+1}+\beta f_m)\beta+2\alpha(-\rho_{m+1}-\alpha f_m)\right]\nonumber\\
& &\qquad=2\epsilon(2pA_{m+1}-2qC_{m+1}-\beta\sigma_{m+1}-\alpha\rho_{m+1})\nonumber\\
& &\qquad=2\epsilon {B_{m+1}}_x.
\end{eqnarray}
Based on the obtained results, we are able to establish the following generalized super integrable hierarchy of Dirac type
\begin{eqnarray}\label{b9}
& &\left[\begin{array} {c}
p \cr
q \cr
\alpha \cr
\beta
\end{array}\right]_{t_m}=K_m=\left[\begin{array} {cccc}
2A_{m+1}+4\epsilon qB_{m+1} \cr
-2C_{m+1}-4\epsilon pB_{m+1} \cr
\sigma_{m+1}+2\epsilon\beta B_{m+1} \cr
-\rho_{m+1}-2\epsilon\alpha B_{m+1}
\end{array}\right],
\end{eqnarray}
among which, the nonlinear super system with $m=2$ is given by
\begin{eqnarray*}
\left\{ \begin{array}{l}
p_{t_2}=-\frac{1}{2}q_{xx}+q(p^2+q^2)+2q\alpha\beta-\alpha\alpha_x+\beta\beta_x
-2\epsilon[p(p^2+q^2)_x+p(\alpha\beta)_x+2p_x\alpha\beta-2q\alpha\alpha_x-2q\beta\beta_x]\\
\qquad -2\epsilon^2q(p^2+q^2)(p^2+q^2+4\alpha\beta),\\
q_{t_2}=\frac{1}{2}p_{xx}-p(p^2+q^2)-2p\alpha\beta-\alpha\beta_x+\alpha_x\beta
-2\epsilon[q(p^2+q^2)_x+q(\alpha\beta)_x+2q_x\alpha\beta+2p\alpha\alpha_x+2p\beta\beta_x]\\
\qquad +2\epsilon^2p(p^2+q^2)(p^2+q^2+4\alpha\beta),\\
\alpha_{t_2}=-\beta_{xx}-q\alpha_x+p\beta_x+\frac{1}{2}p_x\beta-\frac{1}{2}q_x\alpha+\frac{1}{2}(p^2+q^2)\beta
-\epsilon[(p^2+q^2)_x\alpha+2(p^2+q^2+2\alpha\beta)\alpha_x\\
\qquad-(p_xq-pq_x)\beta]-\epsilon^2(p^2+q^2)^2\beta,\\
\beta_{t_2}=\alpha_{xx}+q\beta_x+p\alpha_x+\frac{1}{2}p_x\alpha+\frac{1}{2}q_x\beta-\frac{1}{2}(p^2+q^2)\alpha
-\epsilon[(p^2+q^2)_x\beta+2(p^2+q^2+2\alpha\beta)\beta_x\\
\qquad+(p_xq-pq_x)\alpha]+\epsilon^2(p^2+q^2)^2\alpha.\end{array}
\right.\end{eqnarray*}

\section{Conclusions and discussions}
In this letter, we first have proposed a generalized super Dirac matrix spetral problem by adding the known super Dirac matrix spectral problem with a nonlinear term. Then we have established a generalized super integrable hierarchy of Dirac type by the standard procedure. 

In fact, there are some other possibilities for how to choose the nonlinear term $h$ in $U$. One of such examples is to choose $h=\sum\limits_{j=0}^N \epsilon_{j}(p^2+q^2+2\alpha\beta)_{jx}$ which will leads to different super integrable systems. Beside this generalization, there are also multicomponent generalizations of the super Dirac matrix spectral problem. For example, we can take $U$ as
\begin{eqnarray}\label{f2}
& & U=\left[\begin{array} {ccc}
p & \lambda E+q+h &\alpha\cr
-\lambda E+q-h & -p &\beta\cr
\beta &-\alpha &0
\end{array}\right],
\qquad h=\epsilon(p^2+q^2+2
\alpha\beta), \nonumber
\end{eqnarray}
where $E$ is an unit matrix of $n$-th order, \,$p={\rm diag}(p_1, p_2, \cdots, p_N)$,\, $q={\rm diag}(q_1, q_2, \cdots, q_N)$,\, $\alpha={\rm diag}(\alpha_1, \alpha_2, \cdots, \alpha_N)$ and $\beta={\rm diag}(\beta_1, \beta_2, \cdots, \beta_N)$.
We hope more results can be presented in this direction in the future work.

\section*{Acknowledgements}


This work is supported by the National Natural Science Foundations of China (11271266, 11371323).

\vspace{0.3cm}

\end{document}